\documentclass[]{elsart}
\usepackage{epsfig}
\begin{document}
\begin{frontmatter}
\title{Front-End electronics and integration of ATLAS pixel modules}

\author{F. H\"ugging\thanksref{author}} \\ {on behalf of  the ATLAS Pixel Collaboration\cite{tdr}}
\address{Universit\"at Bonn, Physikalisches Institut \\ Nu\ss allee 12, D-53115 Bonn, Germany}
\thanks[author]{phone: +49(228)733210, e-mail:
huegging@physik.uni-bonn.de}
%\thanksref{bmbf}
%\thanks[bmbf]{work supported by BMBF under contract 05H8PEA1}

\begin{abstract}
For the ATLAS Pixel Detector fast readout electronics has been
successfully developed and tested. Main attention was given to the
ability to detect small charges in the order of $5,000~e^{-}$
within $25~$ns in the harsh radiation environment of LHC together
with the challenge to cope with the huge amount of data generated
by the 80 millions channels of the Pixel detector. For the
integration of the $50~\mu$m pitch hybrid pixel detector reliable
bump bonding techniques using either lead-tin or indium bumps has
been developed and has been successfully tested for large scale
production.
\end{abstract}
\end{frontmatter}

\section{Introduction}
\label{sec:intro} The Pixel Detector of the ATLAS experiment at
the LHC, which is currently under construction at CERN, Geneva, is
the crucial part for good secondary vertex resolution and high
b-tagging capability. Therefore high spatial resolution and fast
read-out in a multi-hit environment is needed and leads to the use
of a hybrid silicon pixel detector. This detector will consist of
three barrel layers and three disks in each forward direction to
have at least three space points up to $|\eta | = 2.5$ for
particles with a transversal momentum greater than $500$
MeV~\cite{tdr}.

The smallest detector unit will be a module made of one silicon
pixel sensor and sixteen readout chips connected with high density
bump bonding techniques to the sensor. Each module consists of
about $50,000$ pixel cells with the size of $50\mu  m \cdot 400\mu
m$ to reach the required spatial resolution of $12\mu m$ in
$r\phi$-direction. The main requirement for the ATLAS pixel
modules is high radiation tolerance of all components in a harsch
radiation environment close to the interaction point. The
integrated design fluence is $10^{15}cm^{-2}$ 1 MeV neutron
equivalent coming predominantly from charged hadrons over ten
years for the outer barrel layers and disks and five years for the
innermost layer. Secondly the total need of about $1,800$ modules
requires very good testability of all components before assembly
and high fault tolerance in long term operation as they are not
supposed to be exchanged during the whole foreseen ten years
lifetime of the ATLAS experiment. The material budget for the
whole pixel detector is very strict to affect the later detector
parts as less as possible. This leads to a total amount of
material less than $1.2~\%$ of one radiation length per module not
including further services and support structures.

\section{Module concept}
\label{sec:concept} 
\begin{figure}[htb]
\begin{center}
\epsfig{file=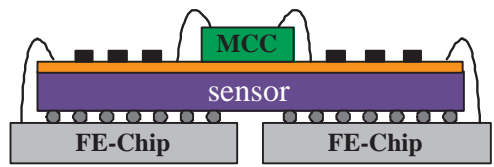,width=0.5\linewidth} \caption{Cross-section
of an ATLAS pixel module.} \label{fig:x-sec}
\end{center}
\end{figure}

A cross-section of an ATLAS pixel module can
be seen in figure~\ref{fig:x-sec}. The module basically consists
of a so called bare module which meant the sixteen readout chips
bump bonded to the silicon pixel sensor. The size of the module is
roughly given by the size of the sensor $18.6\cdot 63.0~$mm$^2$.
The readout chips with are placed in two rows of eight. The
front-end chips are slightly longer than the sensor width to be
able to reach the wire-bond pads of the chips. The interconnection
techniques using fine pitch bump bonding is either done with Pb/Sn
by IZM\footnote{Institut f\"ur Zuverl\"assigkeit und
Mikrointegration, Berlin, Germany.} or with Indium by
AMS\footnote{Alenia Marconi Systems, Roma, Italy.}. The total
active area of one module is $16.4\cdot 60.8~$mm$^2$ with $46080$
pixel.

For routing the signals to the MCC~\cite{mcc} and further to
off-module electronics and providing the analog and digital supply
powers to the readout chips a two layer high density interconnect
(flexible capton cable) is used. Also the flexible hybrid carries
the MCC chip which performs control and clock operations. It is
glued to backside of the sensor and has roughly the size of the
sensor tile. To connect the flex hybrid with the Front-End chips
and with the MCC standard wire bond techniques are used;
altogether around 350 wire-bonds are needed for a complete module.
The module is glued chips down to the mechanical support which
also serves as cooling structure. The whole power consumption of
about $4~$W per module has to be dissipated through the readout
chips to allow an operation of the module at $-6^{\circ}$C.

\section{Front-End electronics}
\label{sec:fe} Each FE-chip contains of 2880 pixel cells arranged
in 18 columns times 160 rows. Every pixel cell consists a charge
sensitive amplifier, a discriminator for zero suppression and a
digital readout logic to transport the hit information to the
periphery. The signal charge can be determined by measuring the
width of the discriminator output signal. Hits are temporarily
stored in one of the 64 buffers per column pair located at the
chip periphery until a trigger signal selects them for readout.
Hits with no corresponding trigger signal are discarded after the
programmable trigger latency of up to 256 clock cycles of 25 ns.
All hits corresponding to trigger are sent through one serial LVDS
link to module controller chip which builds full modules events
containing hit information of all 16 FE chips on a module.

\begin{figure*}[htb]
\begin{center}
\epsfig{file=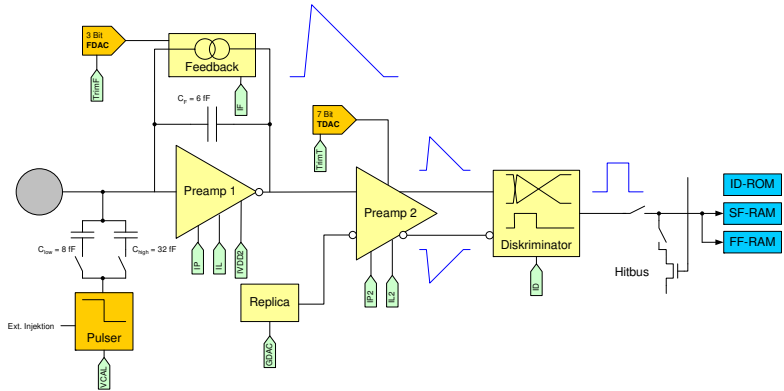,width=0.8\linewidth} \caption{Block
diagram of the pixel unit cell. The output signal of a charge
sensitive amplifier is compared to a threshold voltage to detect
hits. The time of the rising and falling edge of the discriminator
is recorded.} \label{fig:pixel}
\end{center}
\end{figure*}

Figure~\ref{fig:pixel} shows a block diagram of most important
elements in the pixel cell. Charge deposited in the sensor is
brought to the pixel through the bump bond pad which connects to
the input of a charge sensitive amplifier. An inverting folded
cascode amplifier is fed back by a capacitor of $C_{f}=6.5~$fF
integrated into the lower metal layers of the bump bond pad. The
feedback capacitor is discharged by a constant current so that a
triangular pulse shape is obtained. As consequence of this
particular pulse shape, the width of the discriminator output
signal is nearly proportional to the deposited charge. This signal
is called 'time over threshold' or ToT, is measured in units of
the bunch crossing clocks (25 ns) and provides an analogue
information of evry hit with a resolution of 4-6 bit. The feedback
current $I_{f}$ is set globally with an 8 bit on-chip bias DAC so
that a compromise between a good ToT resolution and small dead
time in the pixel can be found. The constant feedback current
generation is part of a circuit which compensates for detector
leakage current. This must be sourced by the electronics because
the pixel sensor is dc-coupled to readout chip. The compensation
circuit can cope with more than 100 nA detector leakage current
per pixel.

The threshold generation of each discriminator is done within
every single pixel. A small threshold dispersion between the 2880
pixel is crucial in order to achieve low threshold settings
without an increased noise hit rate. Therefore a number of 7
individually programmable trim bits in each pixel covers a wide
range of thresholds between $1,000~e^{-}$ and $10,000~e^{-}$.
These local trim bits affects both side of a differential pair
amplifier leading to a linear threshold vs. DAC behaviour in the
central part of the covered threshold range. Furthermore a
globally set 5 bit DAC, so called GDAC, allows to move the
threshold for the whole chip without losing the triming between
the individual pixel.

Every pixel contains a calibration circuit used to inject known
charges into the input node. This circuit generates a voltage step
which is connected to one of two injection capacitors for
different charge ranges.

The output signal of the discriminator in every pixel can be
disabled with a local mask bit. A fast OR of a programmable subset
of pixel is available for testing purposes and for self-triggered
operation of the chip by using this OR signal after delaying it as
a trigger. The 14 configuration bits used in every pixel are
stored in SEU tolerant DICE cells~\cite{dice} which can be written
and read back by means of a shift register running vertically
through the column. The layout of this cell has been significantly
changed with respect to the previous chip (FE-I1) in order to
improve the SEU tolerance. For the globally set DACs a basic
triple redundancy scheme with majority readback is used.

The basic philosophy of the readout is to associate the hit to
unique bunch crossing by recording a time stamp when the rising
edge of the discriminator occurs. The time of the falling edge is
memorized as well in order to calculate the ToT as the difference
of the two values in units of the bunch crossing clock. The two 8
bit RAM cells used for this purpose are classical static memory
cells.

The readout scheme can be divided into four elementary tasks
running on the chip in parallel. First time stamps generated by an
8 bit Gray counter and clocked with bunch crossing clock are
stored for the rising and falling edge of the discriminators in
the pixel. After receiving the falling edge a hit flag is set and
the pixel is ready to be processed. Secondly as soon as hits are
flagged to the Column Control Logic the uppermost hit pixel is
requested to send its rising and falling edge time stamps together
with its ID down the column where is information is stored in a
free location of the End-of-Column (EoC) buffer pool. The pixel is
then cleared and the scan continues the search for other hit
pixel. Thirdly the hit stays in the EoC buffer until the trigger
latency has elapsed. The leading edge time stamp of the hits in
the buffer are therefore permanently compared to the actual time
stamp minus the fix but programmable latency. When the comparison
is true and a trigger signal is present the hit is flagged as
valid for readout, it is discarded otherwise. Incoming triggers
are counted on the chip, the trigger number is stored together
with the flag in the EoC buffer so that several trigger do not
lead to confusion. A list of pending triggers is kept in a FIFO.
Lastly a Readout Controller initiates the serial readout of hit
data as soon as pending triggers are present in the Trigger FIFO,
the EoC buffers are searched for valid data with correct trigger
numbers. The column and row address and the ToT of these hits are
serialized and send to the MCC. The Readout Controller adds a
start-of-event and an end-of-event word together with error and
status bits to the data stream.

The recent chip (FE-I2) has been designed in a quarter micron
technology with 6 metal layers using radiation tolerant layout
rules~\cite{snoeys}. Tests of the first batches received back from
the vendor from May 2003 on showed that the design is fully
functional except one problem dealing with race condition in the
control block leading to the need to operate the chip with a
reduced digital supply voltage of 1.6 V instead of the nominal 2.0
V. But this problem has been fixed via a re-distribution of the
clock signal by changing 2 metal layers slightly in the back-end
processing (FE-I2.1).

\section{Results of prototype modules}
\label{sec:results}

\begin{figure}[htb]
\begin{center}
\epsfig{file=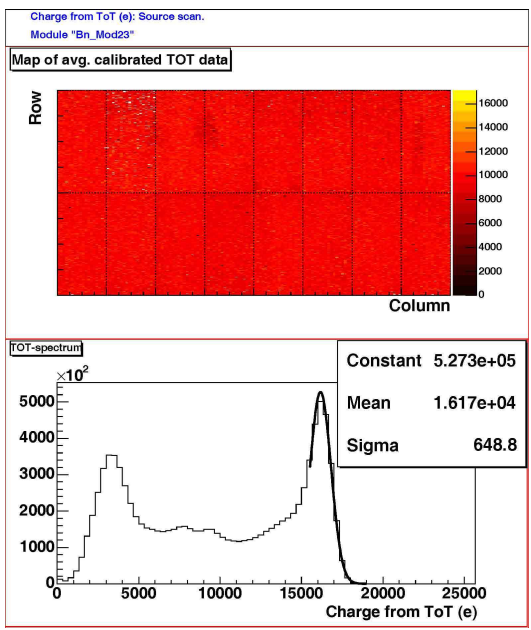,width=0.5\linewidth}
\caption{Am$^{241}$-Spectrum measured with an ATLAS pixel module
using the self-trigger capabilities and the ToT charge
information. Each channel of the module has been individually
calibrated and the shown spectrum is a sum over all pixel without
any clustering.} \label{fig:source}
\end{center}
\end{figure}

Several ten modules have been built with older chip generation
(FE-I1) in order to qualify the bumping process and the whole
module production chain. As an example of the overall good quality
of these modules one can see in figure~\ref{fig:source} the hitmap
and spectrum obtained with one module illuminated by an
Am$^{241}$-source from the top using the self-trigger possibility
of the chip. Only less than 10 out of $46,080$ pixel don't see
source hits showing the excellent bump quality of this module
which is true for almost all modules. For every channel the ToT
response was calibrated individually using the integrated charge
generation circuit. Afterwards this calibration was applied to the
source measurement in order to obtain the shown spectrum which is
a sum over all pixel. Since no clustering of hit data was applied
the spectrum is relatively broad but one can clearly see the main
$60~$keV photo-peak roughly at the expected value of
$16,600~e^{-}$. More results obtained with FE-I1 modules can be
found in~\cite{jgrosse}.

\begin{figure}[htb]
\begin{center}
\epsfig{file=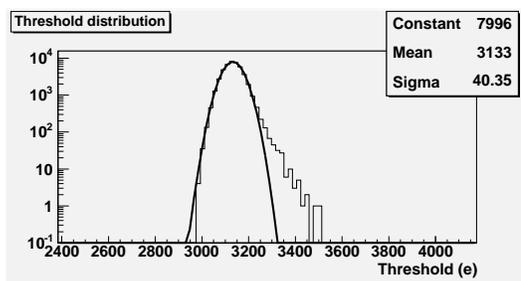,width=0.5\linewidth} \caption{Threshold
distribution of a module built with FE-I2.1 chips after the tuning
procedure.} \label{fig:threshold}
\end{center}
\end{figure}

\begin{figure}[htb]
\unitlength1cm
\begin{minipage}[t]{6.5cm}
\epsfig{file=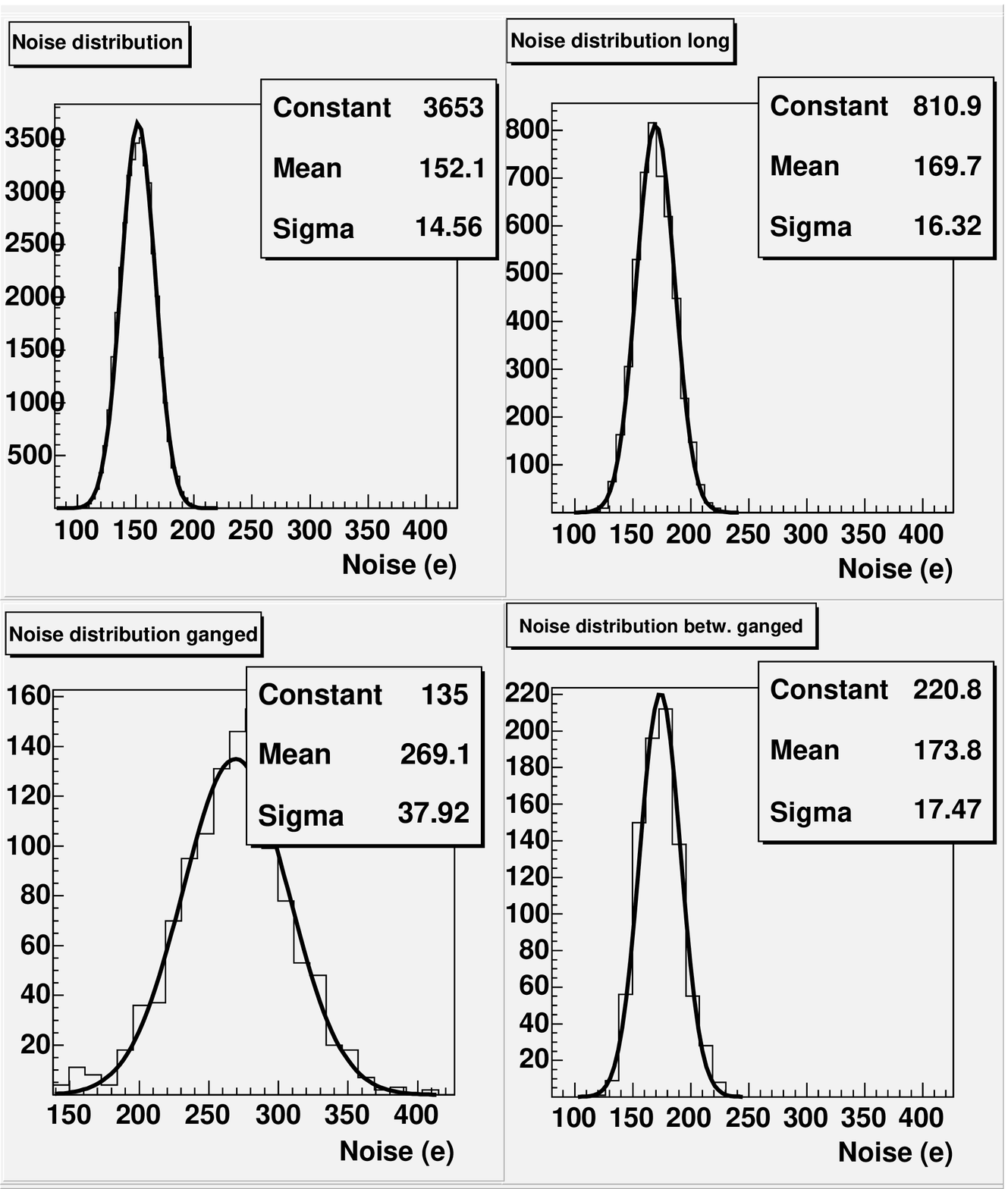,width=\linewidth} \caption{Noise
distributions for the different pixel types of a module built with
FE-I2.1 chips after the tuning procedure.} \label{fig:noise}
\end{minipage}
\hfill
\begin{minipage}[t]{6.5cm}
\epsfig{file=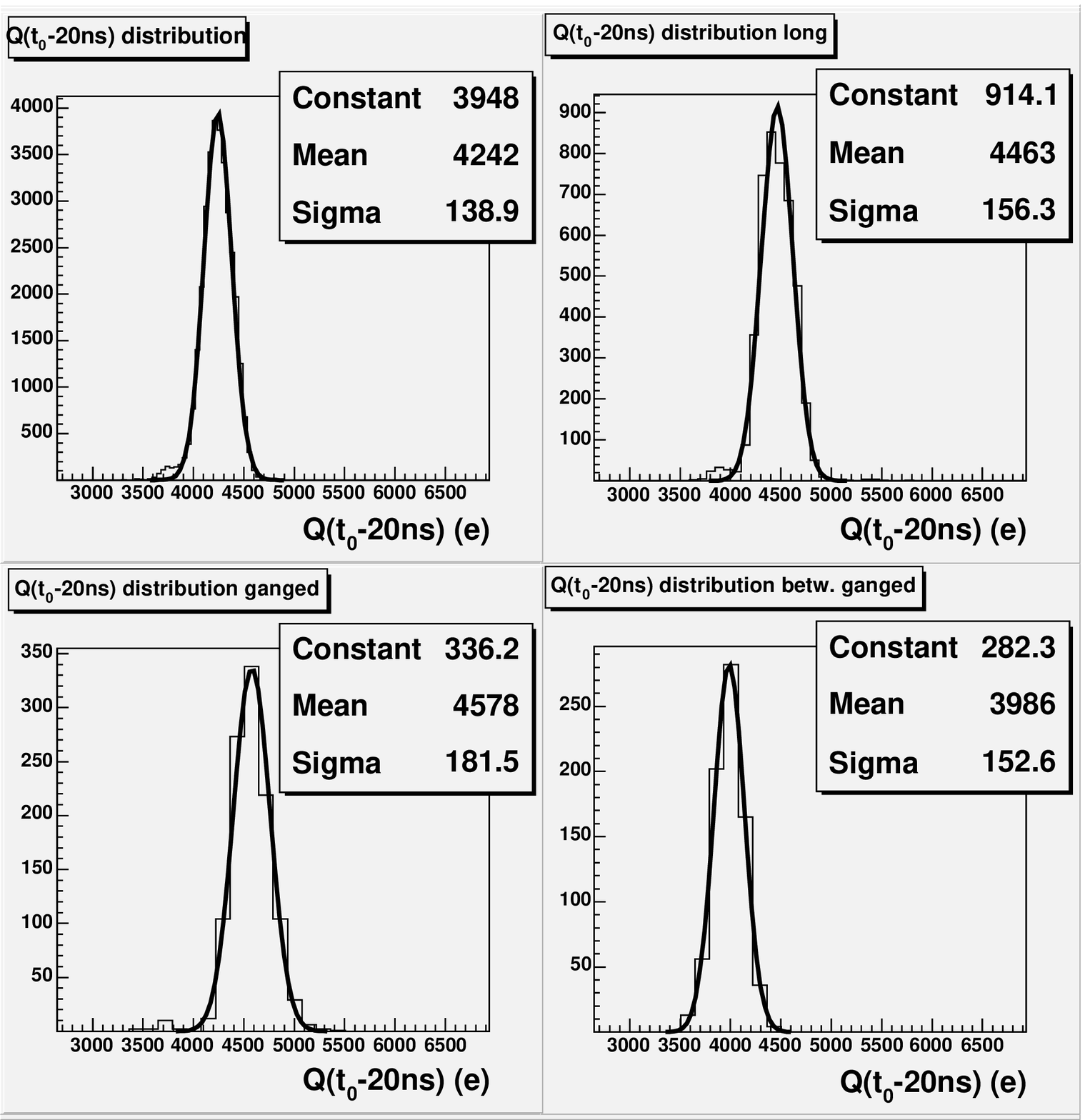,width=\linewidth}
\caption{In-time-threshold distributions for the different pixel
types of a module built with FE-I2.1 chips tuned to an average
threshold of $3,100~e^{-}$} \label{fig:twalk}
\end{minipage}
\end{figure}

Recently first modules with the newer chip generation (FE-I2.1)
has been built. Figure~\ref{fig:threshold} and~\ref{fig:noise}
show the threshold distribution and noise measured for a whole
module after the trimming of the individual pixel to a threshold
of roughly $3,100~e^{-}$. All these measurements were done with
internal charge injection circuit. One can see that the reached
overall threshold dispersion is only $40~e^{-}$ which has to be
compared to value of $100~e^{-}$ reached by the older chip
generation. Note further that there is no single pixel with an
threshold lower than $2,900~e^{-}$. This shows the high tuning
capability of this chip which is important to reach small
thresholds on the whole module without any extra noisy pixel
especially after irradiation.

The measured mean noise of the module is with $150~e^{-}$ for the
standard pixel much better than the specification of $300~e^{-}$,
even the longer and so called ganged pixel located in the
inter-chip regions which show due to the higher input capacity a
higher noise of $170~e^{-}$ and $270~e^{-}$ resp. are well below
this specification.

The most demanding requirement of LHC is the time correct hit
detection within one bunch crossing cycle (25 ns) together with
the strict power budget of less than 4 W for the whole 16 chip
module. Therefore not the real threshold is important but the
threshold taken only hits into account which arrive within in time
slot of 20 ns after the injection. Using a special delay circuit
inside the MCC measurements with the needed accuracy are possible
and in figure~\ref{fig:twalk} the results are shown for the same
module tuned to a threshold of $3,100~e^{-}$. Because the timing
is very sensitive to the input capacity of the amplifier the
in-time-threshold is slightly different for the different kind of
pixel. But overall a in-time-threshold of $4,200~e^{-}$ is reached
and also the special pixel don't show a threshold higher than
$4,600~e^{-}$ meaning that only an extra charge of $1,100~e^{-}$
and $1,500~e^{-}$ resp. is needed to meet the timing requirement
of ATLAS. This results are very much improved with respect to the
previous chip generation where the in-time-threshold was in the
order of $5,000$-$6,000~e^{-}$ by a tuned threshold of
$3,000~e^{-}$. This exceeds specification of $5,000~e^{-}$ given
by the fact that after irradiation damage only half of the initial
charge of $20,000~e^{-}$ is available.

More extensive studies have been done with these modules,
including beam test measurements and measurements during and after
proton irradiation up to 100 MRad or $2\cdot
10^{15}~$n$_{eq}$cm$^{-2}$ well above the design fluence of 50
MRad and $1\cdot 10^{15}~$n$_{eq}$cm$^{-2}$ respectively. All
these measurement confirmed the good quality of the chip and the
whole module, in particular it was shown that these modules meet
all requirements and are able to operate in the environment of
ATLAS and LHC.

\section{Conclusions}
\label{sec:conclusions} The recent ATLAS pixel Frond-End chip
generation has been successfully produced, assembled to full size
modules and tested. Improvements concerning the timing behaviour
and chip threshold tuning capability with respect to the previous
chip generation has been obtained. All further results including
beam tests and irradiation confirmed the overall good quality of
the chip reaching all requirements. By building more than 100
modules with high quality a lot of experience has been gained with
the fine pitch bump bonding process for pixel module using two
different vendors and technologies showing that the target of
building $2,000$ modules needed for ATLAS in the next two years is
feasible.

\end{document}